# Global Simulation of the Solar Wind: A Comparison With Parker Solar Probe Observations During 2018-2022

Chin-Chun Wu[1], Kan Liou[2], Brian E. Wood[1], and Y. M. Wang[1]

[1] US Naval Research Laboratory, Washington D. C. , USA
[2] Johns Hopkins University Applied Physics Laboratory, Laurel, Maryland, USA

**Abstract**

Global magnetohydrodynamic (MHD) models play an important role in the infrastructure of space weather forecasting. Validating such models commonly utilizes in situ solar wind measurements made near the Earth's orbit. The purpose of this study is to test the performance of G3DMHD (a data driven, time-dependent, 3-D MHD model of the solar wind) with Parker Solar Probe (PSP) measurements. Since its launch in August 2018, PSP has traversed the inner heliosphere at different radial distances sunward of the Earth (the closest approach ~13.3 solar radii), thus providing a good opportunity to study evolution of the solar wind and to validate heliospheric models of the solar wind. The G3DMHD model simulation is driven by a sequence of maps of photospheric field extrapolated to the assumed source surface (2.5 $R_\odot$) using the potential field model from 2018 to 2022, which covers the first 15 PSP orbits. The Pearson correlation coefficient (cc) and the mean absolute squared error (MASE) are used as the metrics to evaluate the model performance. It is found that the model performs better for both magnetic intensity (cc = 0.75; MASE = 0.60) and the solar wind density (cc = 0.73; MASE = 0.50) than for the solar wind speed (cc = 0.15; MASE = 1.29) and temperature (cc = 0.28; MASE = 1.14). This is due primarily to lack of accurate boundary conditions. The well-known underestimate of the magnetic field in solar minimum years is also present. Assuming that the radial magnetic field becomes uniformly distributed with latitude at or below 18 $R_\odot$ (the inner boundary of the computation domain), the agreement in the magnetic intensity significantly improves (cc = 0.83; MASE = 0.49).

# 1. Introduction

The long-term and continuous observations of the solar wind near the Earth's orbit have promoted and contributed to the studies of the solar wind and solar events in this area and provided the "ground truth" for validating global simulation models. On the other hand, anywhere outside this region is less studied and less known because of infrequent visits.

A few global three-dimensional (3-D), time-dependent magnetohydrodynamic (MHD) simulation models are currently available for simulating the solar wind and its evolution globally (e.g., Han, 1977; Han et al. 1988; Hayashi, 2005; Manchester et. al. 2004, Fry et at. 2001; Odstrcil *et al.* 2005, Wu *et al.* 2007a,b; 2020a,b; Pomoell & Poetds, 2018). But only a few of them are geared toward simulating the solar wind in real time. For example, the ENLIL model (Odstrcil *et al.* 2005) is adopted by a number of space weather agencies such as US NASA/CCMC, NOAA, and UK Met Office. ENLIL has been validated by comparing seven Carrington rotations (CR2056-CR2062) background solar wind with observation from Ulysses spacecraft while it was orbiting near-Earth to middle to high latitudes during late declining phase of solar cycle 23 (Jian et al. 2015, 2016). The "European heliospheric forecasting information asset" (EUHFORIA) model is developed (Pomoell and Poetds, 2018) and used by the European scientific community. EUHFORIA has been used to simulate the evolution of a coronal mass ejection (CME) from the Sun to the Earth (e.g., Maharana *et al.* 2023). The G3DMHD model (Wu *et al.* 2020a,b) is less known but has its legacy dating back to the first time dependent, global solar wind model developed by Han *et al.* (1987). The Han model had been used to study several different solar wind structures, including the interplanetary magnetic field (IMF) affected by a solar disturbance, i.e., a pressure pulse (Dryer et al., 1997; Wu *et al.* 1996, 2005) and effects of the background solar wind speed on the propagation of interplanetary shocks (Wu *et al.* 2005). G3DMHD is aimed toward reconstructing the realistic solar wind and CMEs in the inner heliosphere (e.g., Wood et al., 2011, 2012; Wu *et al.* 2007a,b, 2011, 2012, 2016, 2019, 2020a,b, 2022).

The G3DMHD model has also been used, in conjunction with a pressure pulse model at the inner boundary, to study the evolution of the extreme fast (>2500 km/s) CME event that occurred on July 23, 2012 (Liou *et al.* 2014, Wu *et al.* 2022), the "Halloween CME event" in October-November, 2003 (Wu *et al.* 2007a; 2012), multiple CME events (e.g., Wu *et al.,* 2007b, 2012, 2019, 2022) , and interaction between CMEs (Wu *et al.* 2007b, 2012, 2019, 2020a, 2022).

While the G3DMHD model was developed with the goal to be able to reconstruct and predict solar wind parameters at ~1 AU (e.g., Wu et al., 2020a,b) rigorous validation of the model within 1 AU, especially in the region close to the Sun, has not been performed. The NASA Parker Solar Probe (PSP) mission was launched in August 2018 and explored the inner heliosphere at different radial distances sunward of the Earth (the closest approach so far of ~13.3 $R$ $R_\odot$). In this investigation we will take advantage of the solar wind measurements provided by the Parker Solar Probe (PSP). Currently, PSP has swung by the Sun more than 18 times and provided invaluable data to date for the study of the evolution of the solar wind. The solar wind data acquired by PSP also provides an unprecedented opportunity to test the performance of current global simulation models from the critical point to ~1 AU. With this in mind, we have performed global 3-D MHD simulations of the solar wind continuously from October 2018 to December 2022 using G3DMHD.

This paper is organized as follows. In Section 2 we describe the G3DMHD model, including the input setup at the inner boundary and the simulation domain. Simulation results, which includes the comparison for simulation versus PSP observation, are presented in Section 3. Discussion and Conclusions are given in Sections 4 and 5, respectively.

## 2. Global three-dimensional magnetohydrodynamic simulation model (G3DMHD)

The G3DMHD model is a well established data-driven, global time-dependent, 3-D MHD numerical simulation model (Wu *et al.* 2020a,b). It is based on the Han model (Han, 1977; Han *et al*. 1988). The Han code is a fully 3-D, time-dependent, MHD simulation model which was designed to simulate ideal solar wind structures for the region beyond 18 $R_\odot$ (*e.g*., S.T. Wu *et al.* 1993; Wu *et al.* 1996, 1998, 2005). Significant improvements have been carried out over the years, which include expanding the computation domain from a cone (±45°) to the entire globe, replacing the static non-realistic boundary condition with the HAF code (Fry *et al.* 2001) and/or with the synoptic magnetic map-driven boundary condition (Wu et al. 2007a,b), and implementing the message passing interface for more efficient computing (Wu et al., 2019, 2020a,b). The current version of G3DMHD is capable of performing continuous, long-term simulation of the solar wind. This is achieved by feeding the model input at the inner boundary (18 $R_\odot$) with continuous solar synoptic maps.

The G3DMHD model solves a set of ideal-MHD equations using an extension scheme of the two-step Lax-Wendroff finite-difference methods (Lax and Wendroff, 1960). The ideal-MHD equations (in the solar rotation frame) consist of conservation laws (mass, momentum, and energy) as shown in Eqs. (1)–(3) with the induction equation (Eq. 4) to take into account the nonlinear interaction between plasma flow and magnetic field.

$$\frac{D\rho}{Dt} + \rho \nabla \cdot V = 0 \qquad (1)$$

$$\rho \frac{DV}{Dt} = -\nabla p + \frac{\nabla \times B \times B}{\mu_0} - \rho \frac{GM_s(r)}{r^2} r \qquad (2)$$

$$\frac{\partial}{\partial t}\left(\rho g + \frac{1}{2}\rho|V|^2 + \frac{|B|^2}{2\mu_0}\right) + \nabla \cdot \left(V\left\{\rho e + \frac{1}{2}\rho|V|^2 + p\right\} + \frac{B \times (V \times B)}{\mu_0}\right) = -V \cdot \rho \frac{GM_s(r)}{r^2} r \qquad (3)$$

$$\frac{\partial B}{\partial t} = \nabla \times (V \times B) \qquad (4)$$

where *t, r, ρ, V, B, p, e* are time, radius, density, velocity, magnetic field, thermal pressure, and internal energy ($e = p/[(\gamma-1)\rho]$). Additional symbols $\gamma$, $M_\odot$, $G$, $\mu_0$ are the polytropic index, the solar mass, the gravitational constant, and the magnetic permeability in vacuum. $\gamma = 5/3$ is used for this study. The G3DMHD solves the ideal MHD equations (Eqs. 1–4) in the reference frame of solar rotation with the rate of rotation set to 27.3 days per rotation.

### 2.1 Simulation Domain and Setup.

The computational domain for the 3D MHD simulation is a heliocentric spherical coordinate system ($r$, $\theta$, $\phi$) with the Sun-Earth line in the 180° longitude meridional plane. Note that the latitude of the Earth changes with seasons within ±7.2° in this coordinate system, e.g., $\theta = -7.2°$ in early March, $\theta = 0°$ in early June and December, and $\theta = +7.2°$ in early September. The simulation domain covers -87.5° ≤ $\theta$ ≤ 87.5°, 0° ≤ $\phi$ ≤ 360°, and 18 $R_\odot$ ≤ $r$ ≤ 345 $R_\odot$. An open (outlet) boundary condition is adopted at $\theta = ±87.5°$ and $r = 345 R_\odot$, so there are no reflective disturbances. A constant grid size of $\Delta r = 3 R_\odot$, $\Delta\theta = 5°$, and $\Delta\phi = 5°$ is used, which results in (110×36×72) grids. Note that a higher resolution is always desirable. The angular resolution used in this study matches the angular resolution of the driving data (e.g., source surface field). There is no significant benefit from having a finer grids in the angular resolution. Although we can reduce the size of radial grids, we have to consider the time it takes to run the simulation. A simulation of this scale (5 years without gaps) is challenging and has not been performed in the past.

## 2.2 Inputs for the simulation

The G3DMHD model is driven by Wilcox Solar Observatory (WSO) solar photospheric maps extrapolated to 2.5 $R_\odot$ and the expansion factor (Wang et al., 1990a,b) using the potential field source surface (PFSS) model (Altschuler and Newkirk, 1969; Schatten *et al.*, 1969; Hoeksema, 1984; Wang and Sheeley, 1992). The radial component of magnetic field ($Br$) at 2.5 $R_\odot$ is extrapolated to 18 $R_\odot$ assuming conservation of the magnetic flux (e.g., $r^2 Br$ = constant). The radial solar wind velocity ($Vr$) is estimated using a modified Wang-Sheeley model (Wu *et al.*, 2020a,b): $Vr$ ($r = 18 R_\odot$) = 150 + 500 $f_s^{-0.4}$ km/s. The longitudinal ($\varphi$) component of the solar wind velocity is determined with the solar rotation at the equator: $V\varphi = 33.8 \times \sin(\theta_c)$ km/s, where $\theta_c$ is the co-latitude and the sine factor approximates the latitudinal distribution. The latitudinal component of the solar wind velocity ($V\theta$) is assumed to be negligible (e.g., $V\theta = 0$). This setup is the same as those used in the ENLIL model (e.g., Odstrcil et al., 2020) except the meridional flow speed is set to zero. The solar wind density is estimated assuming mass conservation (e.g., $\rho = (\rho_o V_o)/V$, where $\rho_o = 2.35\times 10^{-9}$ kg/m$^3$ and $V_o$ takes the average value of $Vr$ at 18 $R_\odot$). The longitudinal and latitudinal components of the magnetic field are derived assuming a corotating field: $B\theta = Br V\theta/Vr = 0$ and $B\varphi = Br V\varphi/Vr$, respectively. Finally, the solar wind temperature is determined assuming conservation of energy, pressure balance, e.g., $2\rho RT + \rho V^2/2 + |B|^2/2\mu_0 + \rho g(r-R_\odot)$ = constant, where $R$ (= 8.314 J/°K–mol) is the gas constant and $g$ (= 274 m/s$^2$) is the solar gravitational acceleration constant. At the solar surface, the magnetic intensity ($B_o$) is a global average value and temperature ($T_o$) = $1.5\times 10^6$ °K.

To simulate the solar wind and its propagation from the inner boundary ($r = 18 R_\odot$) to 1 AU for one Carrington period (~27 days) or less, it requires three synoptic maps for the inner boundary input. The extra maps, one before and one after the current map, remove the discontinuity at the start and end times. This ensures that the simulation reaches the steady state of the solar wind. Stitching of synoptic maps is thus required. Quite often there is discontinuity between them, which may introduce unrealistic disturbances and results in code failures when the disturbances are too large. Smoothing of the maps in longitude in the stitching area is often employed to reduce the discontinuity. For the present simulation, no smoothing is required as the code runs stable across the boundary of the synoptic maps. Note that, only two extra maps are needed for the long-term simulation. A total of 68 (e.g., 2267–2200+1) synoptic maps are used for this study.

Figure 1 shows the time sequence of synoptic maps for $Br$ input at 2.5 $R_\odot$ for the years from 2018 to 2023. The x-axis represents the time (in year) from 2018 to 2023. The y-axis represents the solar latitude (in degree, º). Every five Carrington rotation number is marked on the top of Figure 1. The data resolution is 5º in both longitude and latitude direction for each Carrington map. Each Carrington map lasts for about 27.2753 days. The color bar describes the value of $Br$ with a range of -0.41–0.43 Gauss (marked in the bottom). Figure 1 clearly shows several solar cycle variations of the magnetic field. First, the Sun was quiet during 2018–2019 and started becoming more active in the middle of 2020 (CR2231). Secondly, the solar magnetic field polarity flipped signs in late 2022 (between CR2262 and CR2263). Thirdly, the solar minimum occurred in 2020 since the smallest neutral line ($B_r = 0$, the white solid trace) tilt occurred between January and April 2020 (CR2227–CR2230). After CR2230, the neutral tilt started increasing and reached the south pole at CR2261.

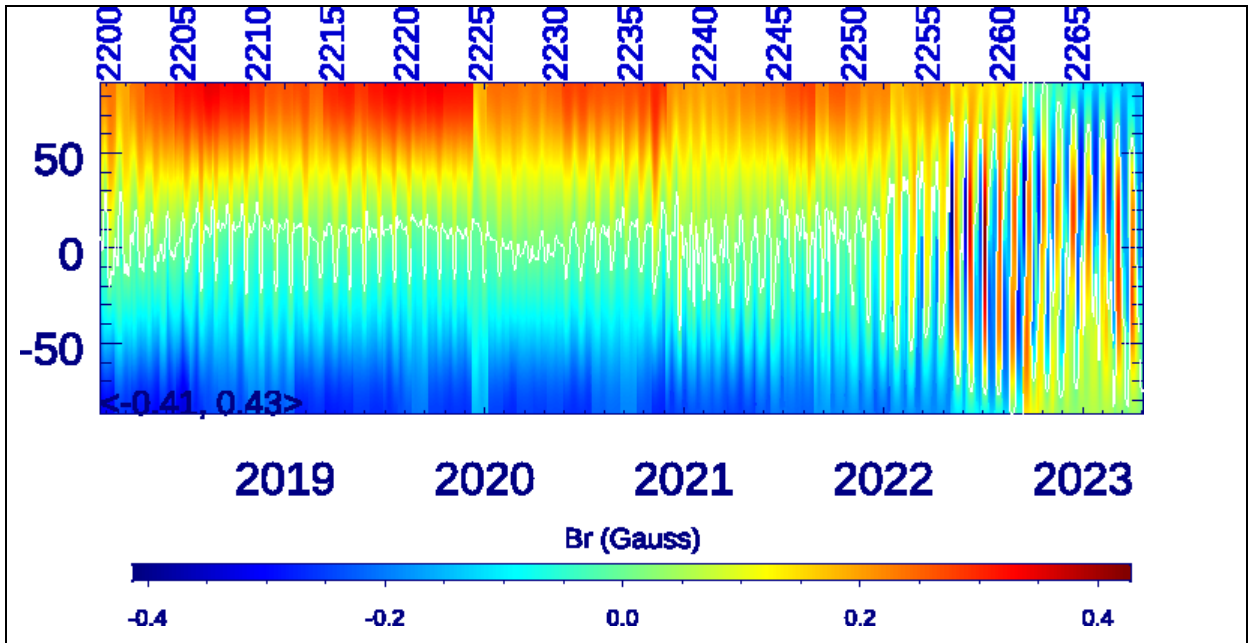

Figure 1. A sequence of synoptic maps at 2.5 $R_\odot$ for 2018 - 2022 derived from the Wilcox Solar Observatory (WSO) photospheric measurements using the potential field source surface model. The Carrington rotation numbers are provided on the top of the Figure. The x-axis represents the time/year of maps and the y-axis represents the solar latitude (in degree). The color bar, which provides the scale of the radial magnetic field ($B_r$), is given at the bottom of the figure. The minimum (= -0.41 Gauss) and maximum (= 0.43 Gauss) values of $B_r$ are listed near the bottom-left corner.

## 3. Simulation Results

We perform G3DMHD simulation using the time sequence of boundary conditions described in Section 2 to construct the global solar wind within 345 $R_\odot$ (1.6 AU). For each time step, the input data of 27.2725 days, which is equivalent to 360° in longitude, is recalculated from the synoptic maps to fill the 72 longitudinal grids. If the numerical grids are not coincided with the beginning of the synoptic map grids (5° or 9.0917 hour wide each), a simple linear interpolation is used. The same stepping method is applied to the inner boundary continuously until all synoptic maps are exhausted. Sometimes it is convenient to stop the simulation at a desired time (e.g., one year for this study) to check the gross result. The simulation can be re-initiated with the solution saved at the last saved time step.

It takes about 15–50 minutes to run a Carrington map (or ~27 days) by using a Linux desktop computer with an AMD Ryzen 9 3950x (3.6 GHz, 16 cores/32 threads) CPU. It takes about 15–50 minutes to run a Carrington map (or ~27 days) on a 16 cores/32 threads CPU (an AMD Ryzen 9 3950x) with an 8-core setup for the message passing interface (MPI). The exact CPU time varies with the frequency of input/output demands. In this study we set the output for one hour approximately to save time and storage space.

### 3.1 Simulation results in the *r–φ* planes

Panels from (a) to (e) in Figure 2 shows selected snapshots of G3DMHD simulated solar wind radial velocity from 2018 to 2022. In each row, there are four maps corresponding to March, June, September, and December at a plane defined by $\theta = 2.5°$N. The exact date and time for each map is marked on the top of each map. The velocity scale is in the range of 250–700 km/s (see the color bar on the bottom of each map) for all maps for easy comparisons. The blue curves on each panel represent the possible location of the HCS. In addition, the 1 AU distance from the center of Sun is marked as a full circle in the *r–φ* plane.

As shown in Figure 2, there is a clear difference in the flow speed among these maps. The flow speed is faster in 2018 and 2019 and is slower in 2020 and 2021. These velocity maps serve as the gross feature of the solar wind from the late descending phase of solar cycle 24 to the early ascending phase of the solar cycle 25, with 2020 for the solar minimum year. During the descending phase of the solar cycle, the solar wind in the equatorial region is often characterized by co-rotating interaction regions (CIRs), where fast flows originating from the coronal holes meet the slow flows originating from the streamers (Burlaga et al., 1978). During the ascending phase of the solar cycle, the increasing occurrence of CMEs is expected to modify the solar wind significantly. Since CMEs are not simulated in this study, possible CME effects are not present in Figure 2. While including CMEs in our simulation can be beneficial to the present study, it is extremely difficult to implement all CME events in the present study.

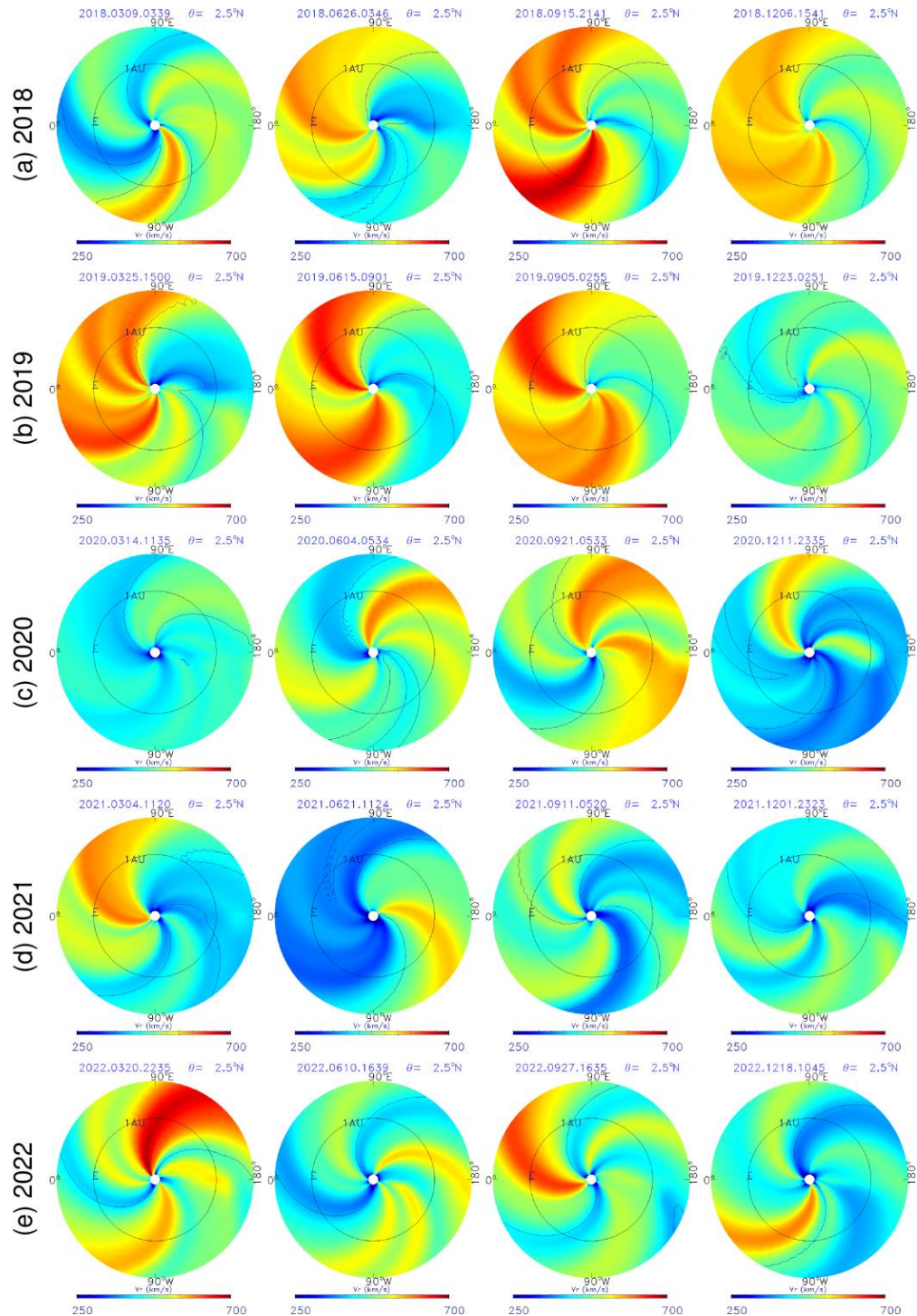

Figure 2. Selected snapshots of G3DMHD simulated solar wind radial velocity in chronicle. Top to bottom in the order from 2018 to 2022. In each row (left to right), there are four maps corresponding to March, June, September, and December at a plane defined by $\theta = 2.5°N$.

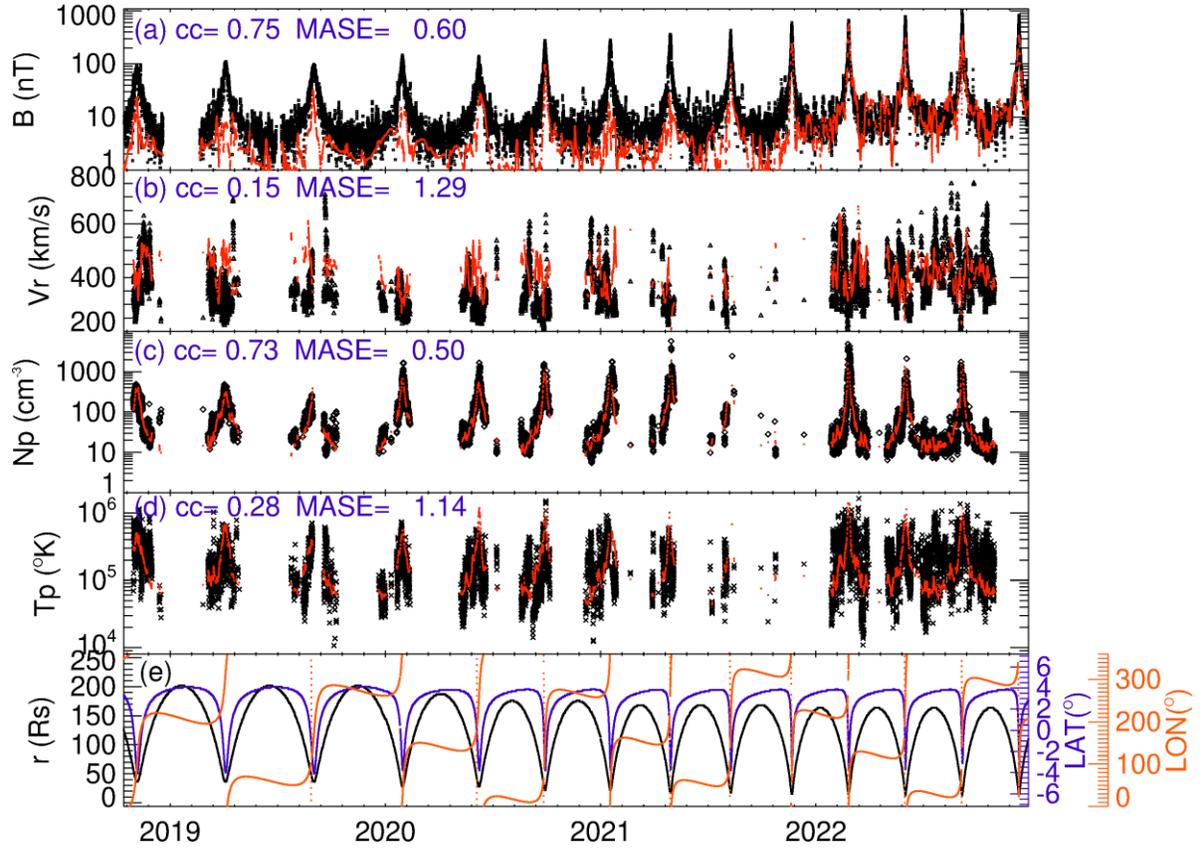

Figure 3. Comparison of PSP in-situ solar wind plasma and magnetic field data verses G3DMHD simulation results during 2018-2022. Panels from top to bottom panels are temperature (units in degrees Kelvin, °K), magnetic field ($B$ in nT), velocity in the radial ($r$) direction ($V_r$ in km/s), number density ($N_p$ in #/cc), and the location of the Parker Solar Probe (PSP) spacecraft (black-, blue-, and orange-lines are the distance from the center of the Sun to the PSP, latitude, and longitude location of the PSP, respectively.

### 3.2 Comparison of simulation results with Parker Solar Probe (PSP) observations

As the main objective of the present study, e.g., comparing the simulation result with in situ solar wind measurements, Figure 3 shows the comparison for hourly resolution of G3DMHD simulation results (red dots) with the hourly PSP (black crosses) solar wind plasma data from the Solar Wind Electrons Alphas and Protons (SWEAP) (Kasper *et al*., 2016) and magnetic field data from the Electromagnetic Fields Investigation (FIELDS) (Bale *et al*., 2016) during 2018–2022. Panels from top to bottom in Figure 3 show the comparison for the (a) magnetic field intensity ($B$ in nT), (b) solar wind radial velocity ($V_r$ in km/s), (c) density ($N_p$ in cm$^{-3}$), and (d) temperature ($T_p$ in °K). The orbit of PSP (black-, blue-, and orange-lines are the distance from the center of the Sun to the PSP, latitudes, and longitudes of the PSP positions, respectively) is shown in Figure 3(e). Note that for the comparison, the simulated solar wind parameters are extracted from the grid where the PSP is located for a given time. The total number of data points for the magnetic field and plasma ($N_p$, $V_r$, and $T_p$) comparison are 28,334 and 9,310, respectively. There are large data gaps in $N_p$, $V_r$, and $T_p$ and these data gaps occur when PSP is near its aphelion, i.e., close to 1 AU.

Table 1. Model validation metrics.

| Year/Parameter | B | | $N_p$ | | $V_r$ | | $T_p$ | |
|---|---|---|---|---|---|---|---|---|
| | cc | MASE | cc | MASE | cc | MASE | cc | MASE |
| 2018 | 0.75 | 1.05 | 0.86 | 0.43 | 0.50 | 0.97 | 0.33 | 1.10 |
| 2019 | 0.67 | 0.85 | 0.73 | 0.62 | -.04 | 2.30 | 0.49 | 1.00 |
| 2020 | 0.82 | 0.78 | 0.78 | 0.45 | 0.01 | 1.83 | 0.34 | 1.35 |
| 2021 | 0.74 | 0.68 | 0.68 | 0.61 | -.06 | 1.78 | 0.24 | 1.26 |
| 2022 | 0.77 | 0.43 | 0.67 | 0.56 | 0.20 | 1.10 | 0.23 | 1.20 |
| 2018-2022 | 0.75 | 0.60 | 0.73 | 0.50 | 0.15 | 1.29 | 0.28 | 1.14 |

cc: Correlation coefficient; MASE: Mean absolute scaled error; N = 28,334 for B and 9,310 for $N_p$, $V_r$, and $T_p$.

Qualitatively the modeled solar wind parameters compare reasonably well with the PSP observations. The values of the Pearson correlation coefficient (cc = $\sum(x_i - \bar{x})(y_i - \bar{y})/\sqrt{\sum(x_i - \bar{x})^2}\sqrt{\sum(y_i - \bar{y})^2}$, where $x_i$ and $y_i$ are the individual sample points and $\bar{x}$ and $\bar{y}$ are the sample mean of variables $x$ and $y$, respectively) are calculated and listed in Table 1. The Pearson correlation coefficient, which ranges between -1 and 1, measures the direction and strength of the tendency to two linearly dependent variables. Therefore, a value closer to 1 means better correlation between the modeled and observed variables. The correlation in $B$ and $N_p$ is especially good. The values of the cc for the entire data is 0.75 for $B$ and 0.73 for $N_p$. However, the correlation for $V_r$ and $T_p$ is poor, cc = 0.15 and 0.28, respectively. Since correlation is a measure of the trend between two variables and is not a measure of the accuracy of forecasts. Here we also calculate the mean absolute scaled error (MASE = $\sum|y_i - \hat{y}_i|/\sum|y_i - \bar{y}|$, where $y$ is the actual value, $\hat{y}$ is the predicted value, and $\bar{y}$ is the mean of the data) (Hyndman and Koehler, 2006) and the results are listed in Table 1. MASE measures the mean of the absolute error of model prediction against the mean of the absolute error from the mean of all data. Therefore, a value of MASE smaller than 1 indicates a better prediction against the mean of the data. Based on this metric, the model performs best in predicting the density (MASE = 0.5), followed by the magnetic field intensity (MASE = 0.6), temperature (MASE = 1.1), and radial velocity (MASE = 1.3). Note that although the correlation for the magnetic field intensity is the highest among the four parameters, there is a clear negative offset or underestimate in the predicted values. In addition, the offset is not random but has a clear trend. It is largest in 2018 and decreases toward 2022 (see Table 1). For the solar wind density, the correlation is the best in 2018 (cc = 0.86) and becomes slightly weaker toward 2022 (cc = 0.67). There is no systematic trend in the values of MASE, which varies in the range of 0.4 and 0.6. For solar wind radial velocity and temperature, there is no systematic trend in the yearly values cc and MASE. There are years (2019–2021) in which correlation for $V_r$ is poor.

It is well known that data smoothing can affect (improve) the values of cc. In this study, we use the hourly solar wind data obtained from Space Physics Data Facility at the NASA/Goddard Space Flight Center to calculate cc and MASE without smoothing the data. In addition, there is a concern about the data gaps to the calculated cc and MASE. The presence of data gaps in the PSP data reduces the "within-series dependence" and thus improves the quality of the calculated cc. For calculating MASE, there are two approaches for estimating the naïve forecast. For continuous time series data, the mean absolute error from the (one-step) prior period are used for the naïve forecast. On the other hand, for non-time series data, the average data value are used as the base predictor (Hyndman and Athanasopoulos, 2018). We applied the latter approach as it is more appropriate to the gappy PSP data.

## 4. Discussion

### 4.1 General discussion

We have performed MHD simulations of the solar wind in the inner heliosphere (18 $R_\odot$ – 345 $R_\odot$) for the period of 2018–2022. The simulated results ($B$, $Vr$, $Np$, and $Tp$) are compared with *in situ* measurements of the solar wind from the Parker Solar Probe (PSP) to evaluate/validate the G3DMHD model. While the comparison is limited to PSP's orbit, the result shows some qualitative agreement. The major finding is that the model performs better in reconstructing solar wind density and magnetic field intensity than in reconstructing the other two parameters ($Vr$ and $Tp$). Solar wind $B$, $Np$, $Tp$ and $Vr$ are in the ranges of 1-1000 (nT), 5-3000 (cm$^{-3}$), $10^4$ -$10^6$ (°K), & 200 – 800 (km/s), respectively. Factor between minimum & maximum are ~1000, 600, $10^2$, and 4 for $B$, $Np$, $Tp$, and $Vr$, respectively. While our simulation results of these four solar wind parameters generally match the observations, the large dynamic range of $B$ may be the major cause for the better cc for $B$.

The generally good agreement between modeled and observed $B$ is expected as it is the only measurements available for the boundary conditions. On the other hand the good agreement in $Np$ and the poor agreement in $V_r$ are totally not unexpected. The boundary value for $Np$ is not measured but derived based on the conservation of mass and an empirical value for the initial mass flux is assigned at the solar surface. Therefore, the value of $Np$ is determined by the value of $V_r$ at the inner boundary. Thus an inaccurate $V_r$ can lead to an inaccurate $Np$. In other words, an accurate $Np$ implies an accurate $V_r$. This is obviously not consistent with the present finding and we do not have an explanation for this result.

The input for $V_r$ is based on the modified WS model (Wu et al., 2020a,b), which is tuned to match in situ observations at 1 AU using 3-D MHD simulations. The poor agreement in $V_r$ may suggest a defect in the empirical formula for the solar wind speed. This defect may be due to the oversimplified formula. Other more sophisticated empirical solar wind speed models such as the Wang-Sheeley-Arge (WSA, Arge et al., 2003) and the coronal hole boundary (DHCB, Riley et al., 2001) models take into account of the shortest distance from the open-close boundary. Some previous works seem to suggest that both WSA and DHCB perform better than the WS model (e.g., Riley et al., 2015). However, Wu et al. (2020b) demonstrated that the modified WS model can outperform the WSA model for solar wind at 1 AU during solar minimum. In addition, the modified WS model is derived based on data acquired in one Carrington rotation (Wu et al., 2020a) and may not be suitable for other solar activity phases. Other models have considered solar activity phases by using different coefficients that are used in their empirical solar wind speed model (e.g., Riley et al., 2015). If the phase of solar cycle is important, one should see a worse result for non-solar minimum years. Indeed, Table 1 clearly shows such a result but only moderately. Obviously, one cannot rule out that the effect of CMEs, which are not modeled in the present study. However, we believe that such an effect should be small as we are able to model $B$ and $N$ reasonably well without including CME events. The solar wind temperature for the boundary is also not measured but derived from the assumption of conservation of energy. This formula depends on the solar wind speed. Thus a poor modeled $T_p$ may be attributed to a poor specification of the solar wind speed at the boundary. However, we cannot rule out the oversimplified solar wind temperature at the solar surface as the cause. A single temperature value is clearly not realistic. Works on improving the inner boundary conditions for the solar wind parameters (e.g., $V_r$ and $T_p$) mentioned above will certainly improve G3DMHD and benefit the community.

Another finding worthy of mention is that the performance of our simulations as indicated by the cc and MASE metrics does not show clear systematic changes over the phase of a solar cycle. This is contrary to our intuition. Our MHD simulations are designed to reconstruct the background solar wind without transit events such as CMEs and shocks. Therefore, it is expected that the best performance would occur in the minimum year of the solar cycle (i.e., December 2019; https://www.weather.gov/news/201509-solar-cycle) because the occurrence of CME events is least frequent. While both cc and MASE values for $N_p$ and $V_r$ in 2018 suggest a better agreement than other periods, we do not see such an effect in $B$ and $T_p$. It is reasonable to argue that a better agreement could be achieved if CME events were excluded in the PSP data and not compared. Unfortunately, classification of the PSP data into CME and non-CME categories is not a trivial task and will not be addressed here. In addition, there are two possible factors affect the comparison result. For example, the ideal MHD model, thus our simulation, is not perfect. PSP often observed large Alfvénic waves and turbulence closer to the Sun. This finding has led to some workers to incorporate MHD turbulence to their simulation models (e.g., Fraternale et al., 2022). Another possible factor is the assigned boundary values are not correct. After all, only one parameter (B in the photosphere) is measured and others are based on some conservation laws. We believe this is the major factor and will be explored in our future work.

### 4.2 Issue of "open flux problem"

In this section, we intend to address the issue of the underestimate of the total magnetic field. According to Figure 3(a), the underestimate decreases with time such that it is largest in 2018 and smallest in 2022. This is in good agreement with some previous results (e.g., Wang and Sheeley, 1988; 1995, Riley et al., 2014, 2019; Jian et al., 2015; Linker et al. 2017; Wallace et al., 2019; Badman et al., 2021) and is referred to as the open flux problem by Linker et al. (2017), although these studies are based on 1 AU data. To further demonstrate the systematic yearly offset in $B$, additional simulation runs are performed by increasing the inputs of the $B_r$ derived from the PFSS model by two, three, and four times at the inner boundary of the simulation domain. Quantitatively, there is little change in the correlation coefficient for $B$, $N_p$, and $T_p$. Although the correlation for $V_r$ becomes slightly weaker for increasing $B$, all coefficient values are too small to be correlated. The MASE values technically remain the same for the three MHD parameters except for the magnetic field intensity.

We may determine the scaling factor for $B$ that best fits the data on a year-by-year basis, and since the magnetic field offset is scalable, one can improve reconstruction of $B$ by scaling up the PFSS magnetic field intensity input at the inner boundary. The smallest MASE value for $B$ are 0.54, 0.67, 0.57, 0.56, and 0.43 for the year 2018 (4$B$), 2019 (3$B$ & 4$B$), 2020 (3$B$), 2021 (2$B$), and 2022 (1$B$), respectively. To avoid the discontinuity between years, a continuous scaling factor, e.g., time-dependent scaling factor, is obtained by fitting a straight line to these yearly scaling factors. The time-dependent scaling factor (or scaling function) is expressed as $SB = 4.63 - 0.767 \times (fyr - 2018)$, where $fyr$ is the fractional year. A simulation run using the scaling factor $SB$ is performed and the result is shown in Figure 4. The results of cc and MASE are listed in Table 2. It is shown that the agreement between the observed and simulated $B$ improves significantly. While the re-scaling of $B$ does not improve the correlation, the overall agreement improves significantly, with the value of MASE reducing to 0.54. While this re-scaling of $B$ improves the performance of our simulation result, one has to be cautious about the use of the scaling function as it has not been tested with more data. For example, the time-dependent scaling factor is derived from data acquired during the ascending phase of current solar cycle. It is clearly not useful for the descending phase as the expected trend in $B$ might require a switch sign for the fitting slope. In conclusion, this type of $B$ re-scaling has no physical ground other than a demonstration that $B$ is scalable.

| Table 2. Model validation metrics for the case with variable $B$ scaling model. | | | | | | | | |
|---|---|---|---|---|---|---|---|---|
| Year/Parameter | $B$ | | $N_p$ | | $V_r$ | | $T_p$ | |
| | cc | MASE | cc | MASE | cc | MASE | cc | MASE |
| 2018 | 0.73 | 0.63 | 0.85 | 0.43 | 0.50 | 1.01 | 0.32 | 1.10 |
| 2019 | 0.71 | 0.62 | 0.73 | 0.62 | -0.04 | 2.41 | 0.48 | 1.07 |
| 2020 | 0.80 | 0.57 | 0.78 | 0.45 | 0.01 | 1.87 | 0.34 | 1.39 |
| 2021 | 0.74 | 0.68 | 0.68 | 0.61 | -0.06 | 1.78 | 0.24 | 1.26 |
| 2022 | 0.77 | 0.43 | 0.67 | 0.56 | 0.22 | 1.10 | 0.23 | 1.20 |
| 2018-2022 | 0.75 | 0.54 | 0.73 | 0.50 | 0.14 | 1.31 | 0.28 | 1.16 |

The systematic IMF $B$ offset suggests the source of properties varying with the solar cycle. An underestimate of $B$ predicted by the PFSS model has been reported previously. Wang and Sheeley (1988) extrapolated magnetic field strength (27-day average) from the WSO potential field to 1

AU and compared it with in situ observations for the period of 1976 to 1986. They found a factor of ~1–2 difference in $B$ when radius of source surface ($R_{SS}$) is either set to 1.6 or 2.5 $R_\odot$. This work is extended to include data from 1968 to 2022 and 8 different synoptic maps recently (Wang *et al.*, 2022). They concluded that the total open fluxes, extrapolated with $R_{SS} = 2.5\ R_\odot$, underestimate the observed radial component of $B$ by factors of ~2–5 for most synoptic maps.

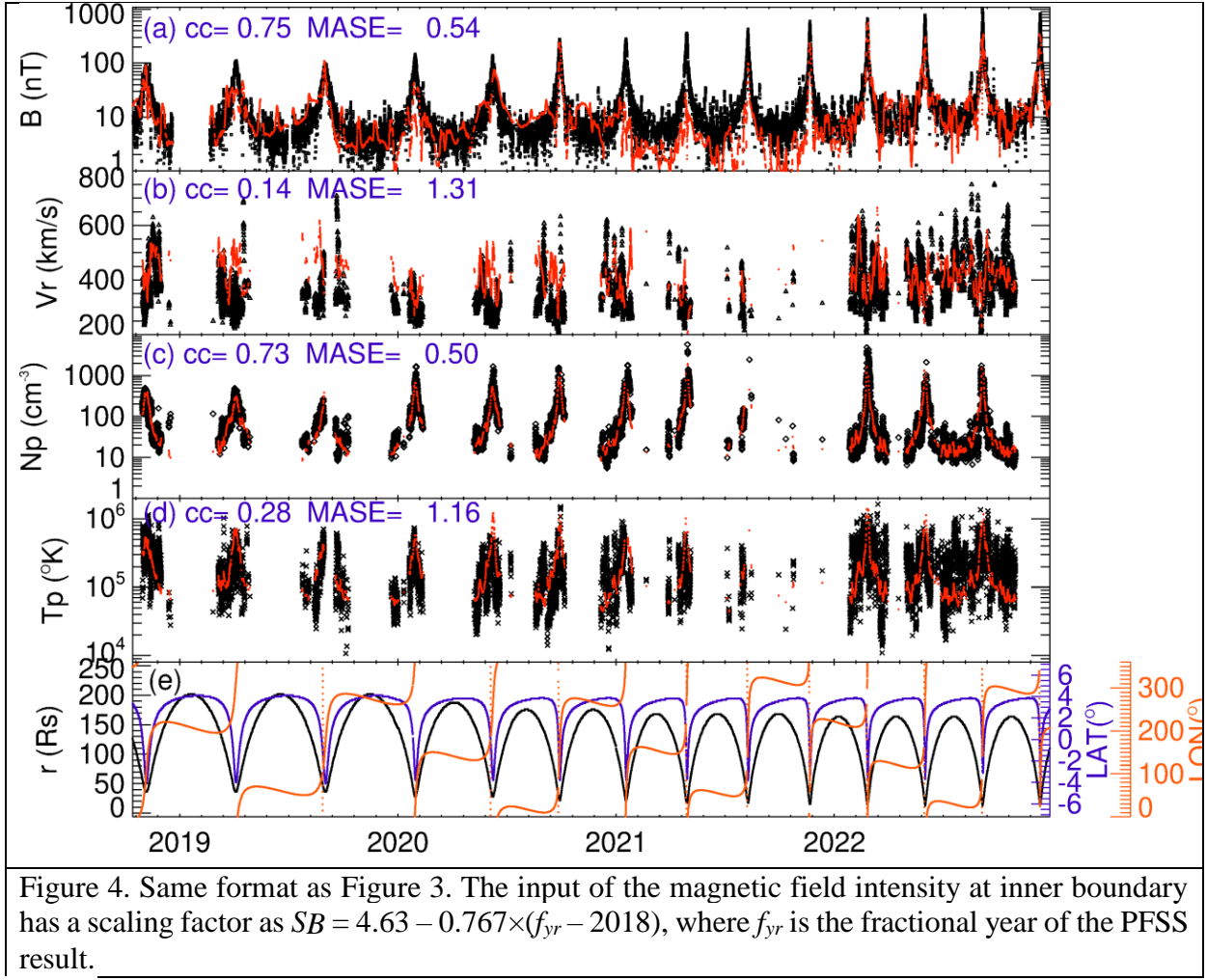

Figure 4. Same format as Figure 3. The input of the magnetic field intensity at inner boundary has a scaling factor as $SB = 4.63 - 0.767 \times (f_{yr} - 2018)$, where $f_{yr}$ is the fractional year of the PFSS result.

According to the magnetic field observations from the Ulysses spacecraft, the total open flux of $Br$ is distributed isotropically at 1 AU (Balogh et al. 1995; Smith & Balogh 2008). The isotropization of the flux could have occurred within $r$ ~10–15 $R_\odot$ due to the zonal flows (e.g., Wang 1996, Zhao and Hoeksema 2010; Cohen 2015). Therefore, for each map we average the magnitude of $Br$ along a fixed longitude and apply the mean value to that longitude at the inner boundary (18 $R_\odot$). This procedure is carried out for all synoptic maps for the entire studied period (e.g., 2018–2022) with no change in the $V_r$ and $N_p$ (note that $T_p$ changes with $B$). Figure 5 shows the simulation result using the latitudinally averaged magnetic maps and the correlation coefficients and MASE values for $B$, $N_p$, $V_r$, and $T_p$ are listed in Table 3. The correlation for $B$ is improved significantly (from cc = 0.75 to cc = 0.83) and the MASE value is reduced by 20% (from 0.60 to 0.49) comparing with simulation results without any modification for $B$. Note that the extremely good correlation (cc = 0.93) for $B$ for the year of 2021 and the value of MASE drops from 0.68 to 0.43 for the year of 2021. Therefore, using constant $|B_r|$ at the inner boundary is a good approach for realistic solar wind simulation.

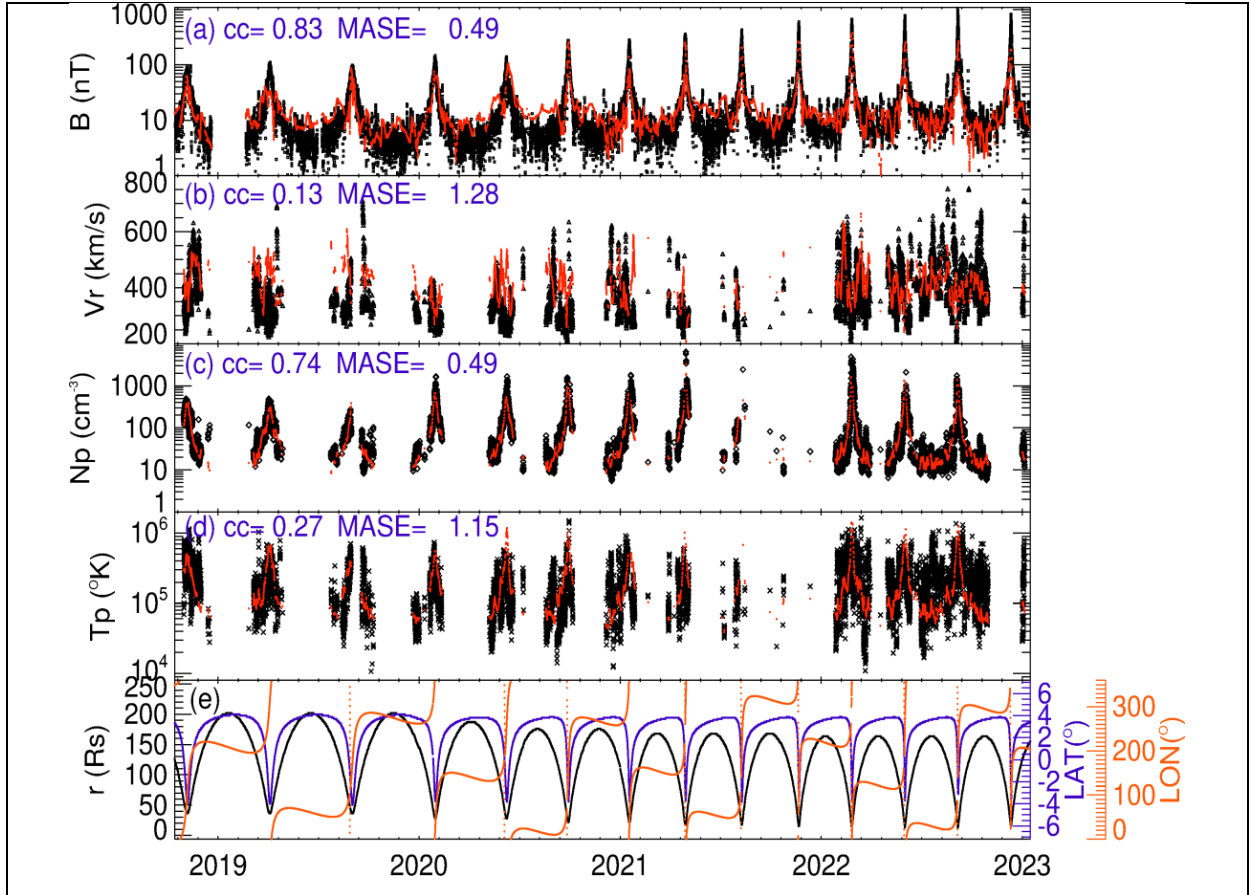

Figure 5. Same format as Figure 3. The input of the magnetic field intensity at inner boundary employs a uniform $B_r$ along a fixed longitude.

Table 3. Model validation metrics for the case with a constant Br at a fixed longitude.

| Year/Parameter | $B$ | | $N_p$ | | $V_r$ | | $T_p$ | |
|---|---|---|---|---|---|---|---|---|
| | cc | MASE | cc | MASE | cc | MASE | cc | MASE |
| 2018 | 0.76 | 0.65 | 0.82 | 0.47 | 0.46 | 1.04 | 0.31 | 1.11 |
| 2019 | 0.84 | 0.51 | 0.73 | 0.64 | -0.06 | 2.42 | 0.47 | 1.14 |
| 2020 | 0.91 | 0.42 | 0.78 | 0.48 | 0.03 | 1.83 | 0.35 | 1.47 |
| 2021 | 0.93 | 0.43 | 0.63 | 0.63 | -0.04 | 2.11 | 0.31 | 1.53 |
| 2022 | 0.83 | 0.51 | 0.65 | 0.57 | 0.15 | 1.10 | 0.22 | 1.21 |
| 2018-2022 | 0.83 | 0.49 | 0.74 | 0.49 | 0.13 | 1.28 | 0.27 | 1.15 |

### 4.3 Radial and longitudinal dependence in G3DMHD performance

Since PSP orbits cover a wide range of the radial distance from the Sun, it would be important to explore if the performance of G3DMHD is radial dependent. We divided PSP's radial component orbit into 3 regions: $r < 0.3$ AU, $0.3$ AU $< r < 0.6$ AU, and $0.6$ AU $< r < 0.9$ AU. The calculated cc and MASE for these 3 regions are listed in Table 4. In general, there is no significant change in the performance G3DMHD in reconstructing the solar wind parameters, e.g., the modeled $B$ and $N_p$ match the observed $B$ and $N_p$ than $V_r$ and $T_p$ do. With regard to the radial dependence, the Table indicates that when there is a good correlation between the modeled and observed parameters ($B$ and $N_p$), the performance of the G3DMHD model reduces with increasing radial distances from the Sun. This result is not surprising because a better agreement between the modeled and observed results closer to the inner boundary is expected if the boundary conditions are correctly prescribed. The values of cc for $B$ and $N_p$ are moderate (0.76 and 0.62, respectively), whereas the values of cc for $V_r$ and $T_p$ are very small (0.09 and -0.05, respectively). Similarly, the values of MASE for $B$ and $N_p$ are only moderate (0.67 and 0.76), whereas the values of cc for $V_r$ and $T_p$ are greater than 1 (1.77 and 1.89). This may suggest that none of the boundary conditions are correctly specified, especially the $V_r$ and $T_p$ parameters. It is also possible that the G3DMHD is not perfect for solar wind modeling. We tend to believe that the former is the major cause because the G3DMHD has been shown to be able to reconstruct solar wind in some cases (e.g., Wu et al. 2016, 2019, 2022).

Note that Table 4 also indicates that the performance of G3DMHD reduces significantly after $r > 0.6$ (cc = 0.18 and MASE = 1.18 for $B$ and cc = 0.17 and MASE = 0.98 for $N_p$). We will show later that the performance of the G3DMHD model also depends on the longitude of observations from the Sun-Earth line.

Table 4. Radial dependence of the G3DMHD model performance.

| $r$-range (AU) | $B$ | | $N_p$ | | $V_r$ | | $T_p$ | |
|---|---|---|---|---|---|---|---|---|
| | cc | MASE | cc | MASE | cc | MASE | cc | MASE |
| < 0.3 | 0.76 | 0.67 | 0.62 | 0.76 | 0.09 | 1.77 | -0.05 | 1.89 |
| 0.3–0.6 | 0.28 | 1.25 | 0.56 | 0.80 | 0.13 | 0.98 | 0.18 | 1.26 |
| 0.6–0.9 | 0.08 | 1.18 | 0.17 | 0.98 | -0.01 | 1.03 | 0.02 | 1.22 |

While the entire photospheric magnetic field is measured "instantaneously", the making of synoptic maps is not. The synoptic map combines the latitudinal stripe of daily magnetic field data near the Sun-Earth line to construct the synoptic maps by assuming the solar magnetic field does not change significantly from one solar rotation to another. Therefore, the reliability of the synoptic maps as the boundary condition for $B$ may be in question because the solar magnetic field in not static. As a result, the performance of G3DMHD should reveal a longitudinal dependence. Here, we divide the longitudinal domain into 12 sectors with 30° for each sector. The values of cc and MASE are calculated for data falling within each sector and are listed in Table 5. The best performance occurs, as expected, when PSP is within 15° (e.g., 345°-15°) from the Sun-Earth line. The values of cc are in the range of 0.64–0.93 for $B$ and 0.45–0.86 for $N_p$. No systematic reduction in cc with the longitudinal separation between the PSP and Sun-Earth line for these two parameters. A modest improvement in the correlation (cc = 0.46) in the modeled solar wind speed occurs near the Sun-Earth line. The result for $T_p$ is consistently poor. We noticed that the cc value is 0.62 at 75°–105°. However, this improvement comes with a large MASE value (3.04). Therefore we do not think this is a good match.

Table 5. Longitudinal dependence of the G3DMHD model performance.

| Longitude angles | $B$ | | $N_p$ | | $V_r$ | | $T_p$ | |
|---|---|---|---|---|---|---|---|---|
| | cc | MASE | cc | MASE | cc | MASE | cc | MASE |
| 345°–15° | 0.93 | 0.38 | 0.78 | 0.47 | 0.46 | 1.40 | -0.06 | 2.09 |
| 15°–45° | 0.85 | 0.56 | 0.86 | 0.38 | 0.06 | 1.23 | 0.10 | 1.10 |
| 45°–75° | 0.78 | 0.48 | 0.76 | 0.46 | 0.42 | 1.26 | 0.10 | 2.05 |
| 75°–105° | 0.64 | 0.75 | 0.57 | 0.62 | 0.05 | 2.96 | 0.62 | 3.04 |
| 105°–135° | 0.88 | 0.54 | 0.70 | 0.57 | 0.29 | 0.94 | 0.11 | 1.31 |
| 135°–165° | 0.86 | 0.44 | 0.45 | 1.10 | 0.45 | 1.07 | -0.32 | 1.73 |
| 165°–195° | 0.75 | 0.69 | 0.51 | 1.00 | 0.25 | 1.60 | 0.18 | 3.93 |
| 195°–225° | 0.86 | 0.55 | 0.85 | 0.61 | 0.36 | 1.00 | 0.39 | 1.00 |
| 225°–255° | 0.82 | 0.52 | 0.81 | 0.63 | 0.13 | 2.48 | -0.26 | 2.45 |
| 255°–285° | 0.84 | 0.52 | 0.78 | 0.43 | 0.36 | 1.14 | 0.09 | 1.25 |
| 285°–315° | 0.88 | 0.56 | 0.73 | 0.60 | 0.23 | 1.22 | 0.02 | 1.20 |
| 315°–345° | 0.87 | 0.40 | 0.85 | 0.96 | 0.00 | 2.16 | -0.17 | 2.04 |

The present study demonstrated that G3DMHD is better suited for reconstruction of the solar wind density and magnetic field intensity. However, it performs poorly in simulating the $V_r$ and $T_p$. It is reasonable to attribute the poor agreement to lack of a realistic input from measurements. All current data driven MHD simulation models are subject to the same vexing drawback. In addition, the poor agreement may be attributed partially to lack of CME simulation. However, the good agreement in molded B and Np with observations and lack of clear solar cycle variation may suggest that effects of CME might be small. Therefore, we believe that the boundary condition must be the major cause. This suggestion is also reinforced by the result of radial and longitudinal analysis as shown in Section 4.3. As part of our future efforts, we will point out a few possible directions to improve the G3DMHD model. First, the simple model of Wang and Sheeley (1990) is used for specifying the solar wind speed at the inner boundary. Although the model has been modified using MHD simulations (Wu et al., 2020a,b), it is for solar quiet times only. This solar wind speed model may be improved by considering other solar conditions. Second, Arge et al. (2003) proposed an improved model (know as the WSA model) of the solar wind that incorporates the angular distance of the field line from the nearest coronal-hole boundary. The WSA-ENLIL model has been adopted by CCMC and has been used in many studies. We will incorporate the WSA model into G3DMHD to test its performance. Thirdly, the solar wind speed and temperature are usually well correlated at 1 AU (Neugebauer and Snyder, 1966; Strong et al., 1966; Burlaga and Ogilvie, 1970; Lopez and Freeman, 1986; Lopez, 1987; Elliot et al. 2010). However, no study was performed sunward of 1 AU. We anticipate that the agreement in $T$ will improve once the agreement in $V$ is improved. On the other hand, the PSP data can be used to test if the $T$-$V$ relationship still hold close to the Sun. This will allow us to build an empirical $T$-$V$ relationship for MHD modeling of the solar wind.

As a final note, the intent of this study is to evaluate the performance of G3DMHD model for the background solar wind. No solar disturbance (e.g., CME) has been applied to the inner boundary. This could be one of the reasons that contribute to some of the poor results. Introducing CME events into solar wind simulation has been performed in some single event studies (e.g., Wu et al. 2016, 2020b, 2022). Modeling multiple CMEs are still rare (e.g., Wu *et al*., 2022). Some studies model the CME with a pressure pulse (e.g., Wood et al., 2011, 2012; Wu *et al.* 2016, 2020b) and some with a flux rope (*e.g*., Manchester *et al*., 2008; Shen *et al*., 2012) or two flux ropes (Koehn *et al.,* 2022). Either approach is still difficult currently without mentioning that it is impractical to include all CME events because identifying CME events also constitutes another difficulty. While the present study suggests that some improvements are required, it encouraged us to implement flux ropes into the G3DMHD model (*e.g*., Wu *et al.* 2024). Once the flux-rope model + G3DMHD model is able to simulate flux-roped CMEs and its driven-shock correctly, we will try to add all available CMEs into the inner boundary and re-run the yearly simulation in the future.

## 5. Conclusions

The main point of this G3DMHD validation work is: (1) The present G3DMHD model is better suited for modeling solar wind density and magnetic field intensity but not for the solar wind speed and temperature within 1 AU. We suspect this is due to the improper boundary conditions adopted by G3DMHD. (2) The well-known underestimate of magnetic field intensity in solar minimum is present in the PSP orbit, it is demonstrated that $B$ is scalable by changing its value at the inner boundary; (3) An improved agreement in the magnetic field intensity can be achieved by assuming an isotropic distribution of $Br$ within 18 $R_\odot$, which also removes the underestimate of magnetic field intensity; (4) Changing the magnetic field intensity at the inner boundary does not lead to significant changes in plasma parameters (e.g., $N_p$, $V$, and $T$), suggesting that the dynamic effect probably dominates the large-scale evolution of solar wind in the inner heliosphere antisunward of 18 $R_\odot$; (5) The performance of G3DMHD reduces in regions away from the inner boundary and the Sun-Earth line. We also point out a few areas of improvements for our future work. Finally, since this is the first effort that a global MHD simulation model is evaluated with the PSP data, the present result can serve as the benchmark for later works.

**Acknowledgment**

All data used in this study are obtained from the public domain. We thank the Parker Solar Probe PI teams and the Space Physics Data Facility at NASA/Goddard Space Flight Center, for providing the hourly solar wind plasma and magnetic-field data. This work was supported partially by the Office of Naval Research (BW, CCW & YMW), and NASA grant of 80HQTR20T0067 (BW & CCW). The work of KL was supported by NRL grant N00173-21-2-C007 to the Johns Hopkins University Applied Physcis Laboratory. The authors thank Dr. Sam Cable from Engility/DoD High Performance Computing Modernization Office PETTT program for his technical assistance in work on parallelizing the G3DMHD code.